\documentstyle[multicol,aps]{revtex}
\input{psfig}
\begin{document}
\newcommand{\la} {\langle}
\newcommand{\ra} {\rangle}
\newcommand{\ep} {\epsilon}
\newcommand{\bu} {\bf u}
\newcommand{\De} {\Delta}
\pagestyle{myheadings}
\markright{submitted to PRL}
\draft
\preprint{Submitted to Phys. Rev. Lett.}
\title{Anomalous Scaling and Structure Instability in 
Three-Dimensional Passive Scalar Turbulence}
\author{Shiyi Chen$^{1,2}$ and Nianzheng Cao$^{1,2}$}
\address{${}^{1}$IBM Research Division, T. J. Watson Research Center,
P.O. Box 218, Yorktown Heights, NY 10598\\
${}^{2}$Theoretical Division and Center for Nonlinear Studies,
Los Alamos National Laboratory, Los Alamos, NM 87545}
\maketitle

\begin{abstract}

The anomalous scaling phenomena of three-dimensional passive scalar 
turbulence are studied using high resolution direct numerical 
simulation. The inertial range scaling exponents of the 
passive scalar increment and the scalar dissipation are obtained. 
The connection between the intermittency structure
and the scaling exponent is examined and the structure instability
of the high amplitude scalar dissipation is used to clarify 
previous experimental results for the scaling exponents. 

\end{abstract}

\begin{multicols}{2}
\narrowtext

The advection of a passive scalar field by a 
turbulent (or stochastic) incompressible velocity field  
is a process that exhibits cascade to small scales.
The dynamics governing a passive scalar is described by the
following equation,
\begin{equation}
\partial T/\partial t + {\bu}\cdot\nabla T = \kappa \nabla^2 T +
{\bf f}.
\end{equation}
Here $T$ is the passive scalar, $\kappa$ is the
kinematic diffusivity, ${\bf f}$ is a random
forcing, and the advective
velocity, ${\bu}$, is governed by the
Navier-Stokes (NS) equations in three dimensions.

The study of the dynamics in the passive scalar system has been 
one of the most active research areas in the field of fluid 
turbulence for the last decade\cite{andy,rhk}. In particular, 
based on a linear ansatz for the dissipation term, 
Kraichnan\cite{rhk} obtains an explicit prediction of the anomalous 
scaling exponents for a class of passive scalar
advected by a white and Gaussian incompressible velocity. 
The research has inspired a large number of papers\cite{pass} over the
past two years, including numerical simulations\cite{rhk1}
and analysis of experiments\cite{ching}.

The passive scalar convected by Navier-Stokes turbulence, however, 
is a more difficult problem where the velocity is far from white and
Gaussian. No inertial range scaling  based on the passive scalar dynamical 
equation (1) has been developed. The local similarity theory 
for the passive scalar, an extension to Kolmogorov 1941 similarity theory 
for the velocity field\cite{k41}, was studied by Obukhov\cite{obu} and 
Corrsin\cite{corrsin}.  For fluid flows at very high 
Reynolds number and with Prandtl number, $Pr = \nu/\kappa$,
close to unity, Obukhov's theory assumes the existence of universal statistics
of fluctuations at so-called inertial-range scales $L \gg r \gg\eta$,
where $\nu$ is the kinematic viscosity, $L$ and $\eta$ are 
the characteristic length scales for the
large scale and the dissipation scale, respectively.
This hypothesis has a series of implications, including 
the 2/3 law for the passive scalar fluctuations in the inertial range:
$\la\De T_r^2\ra\sim\la \ep \ra^{-1/3} \la N \ra r^{2/3}$, 
where $N=\kappa (\nabla T)^2$ is the scalar dissipation function,
$\ep$ is the velocity dissipation function,  $\De T_r
= T(x+r) - T(x)$ is the scalar increment and $\la \cdot \ra$ 
denotes an ensemble average. In general, if there is a scaling
range, the $p$th order structure function, 
$S_p(r) = \la\De T_r^p\ra$, should have a
scaling relation: $S_p(r) \sim r^{z_p}$, with $z_p$ defined as the $p$th
order scaling exponent. 

Taking account of the intermittency correction\cite{kuo} and 
using the refined similarity hypothesis (RSH)\cite{k62,sto}, the 
phenomenological scaling models for the passive scalar, 
including the bi-variate log-normal model by Van Atta\cite{van}, the $\beta$
model\cite{beta} by Frisch {\em et. al.} and the bi-variate
multifractal model by Meneveau {\em et. al.}\cite{meneveau}, lead to
analytical predictions of anomalous scaling exponents.
In a recent note, we have developed a bi-variate log-Poisson 
model\cite{cao-chen}; the resulting scaling exponents agree well with 
experimental measurements. 

The challenge in the real-life experimental measurement is to 
minimize the interference of two probes which measure velocity 
and scalar quantities simultaneously at the same position\cite{meneveau}.
Our research is motivated by the paper by Sreenivasan and 
Antonia\cite{sreeni} who pointed out that two existing experimental
measurements\cite{antonia,meneveau} do not provide a convergent scaling
relation for the passive scalar system. 
In particular, the scaling exponents in \cite{meneveau} were obtained
using the measured dissipation exponents by invoking the 
RSH, showing a saturation for $p \ge 6$.  The scaling exponents from  
another experimental measurement\cite{ruiz} fall between above two results.

Direct numerical simulation of the passive scalar equation (1)
and the Navier-Stokes equations were carried out 
simultaneously. A mesh size of $512^3$ was used 
in a cyclic cubic box for homogeneous isotropic turbulence.
In order to maintain statistical steady states, both the velocity 
field and the passive scalar field were forced for $k < 3$\cite{cao,kerr}. 
The forcing scheme keeps the total energy in the first two shells 
($1\le k < 2$ and $2\le k < 3$) consistent with $k^{-5/3}$. 
In addition, Fourier modes in each shell have equal energy and
the phase of each mode is randomized. 
For simplicity the Prandtl number is fixed to be unity for this study. 
The analysis was carried out for the statistical
steady states at Taylor microscale Reynolds number ${\cal R}_\lambda = 220$.
A spatial averaging over the whole physical domain and a time average over
approximately 10 large-eddy turnover time were used to 
replace the ensemble average. 

\bigskip
\psfig{file=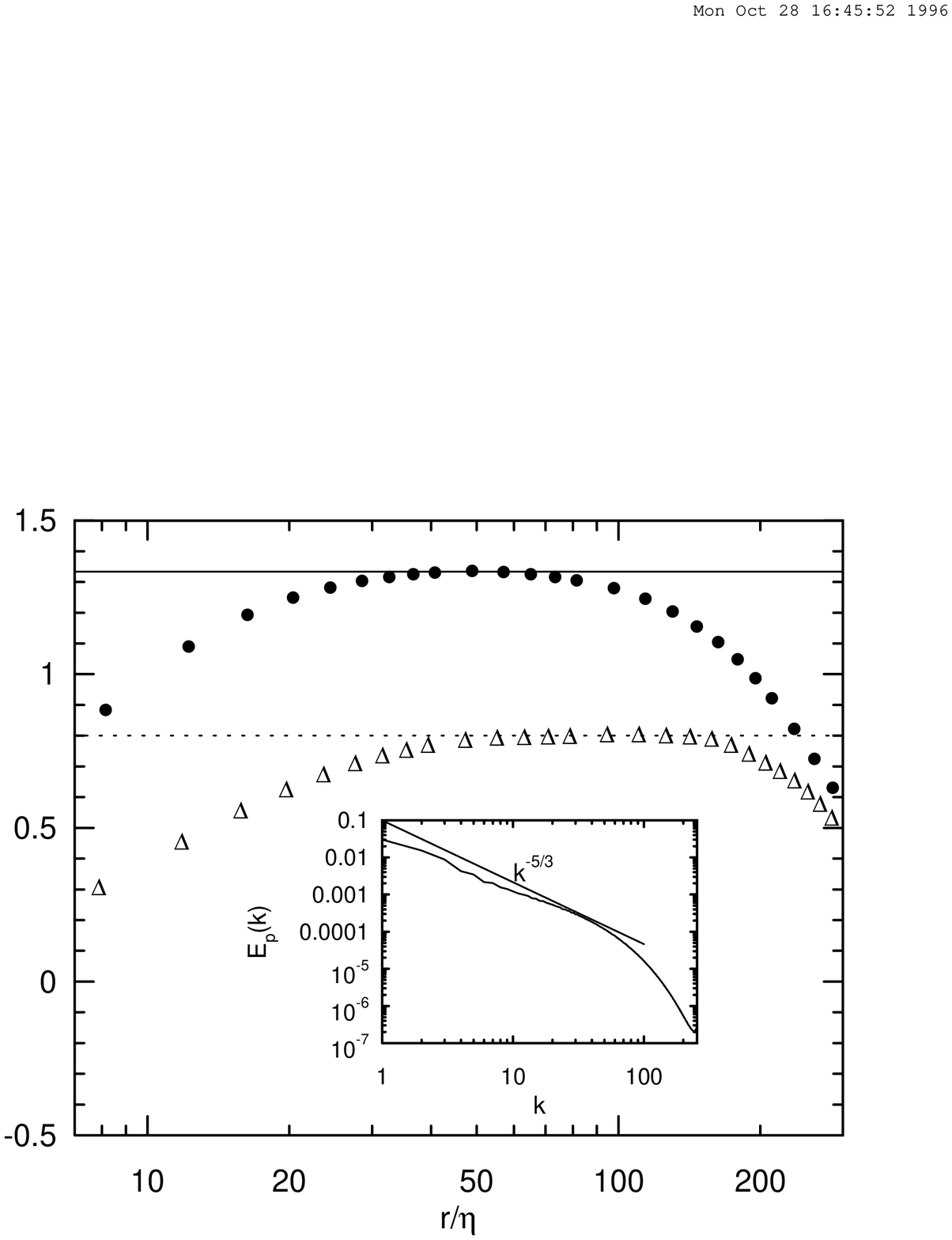,width=220pt}
\noindent
{\small FIG.~1.
Numerical verification of Kolmogorov's 4/5 law and
Yaglom's 4/3 law. The solid line is for Kolmogorov's 4/5 law and
the dashed line is for Yaglom's 4/3 law. The signs are from DNS 
data ($\triangle$ for $-\la (\De u_r)^3 \ra /(\la\ep\ra r)$ and
$\bullet$ for $-\la (\De u_r (\De T_r)^2 \ra /(\la N \ra r)$.
Here $\eta = (\nu^3/\la\ep\ra)^{1/4}$ is the Kolmogorov dissipation length, 
In the inset shows the scalar spectrum compared with Oboukov's $-5/3$ 
law for the inertial range scaling.}
\bigskip

At very high Reynolds number, Kolmogorov's 
4/5 law\cite{k45} is exact in the inertial range:
$\la (\De u_{r})^3 \ra = \la (u(x+r)-u(x))^3 \ra = -4/5\la\ep\ra r $.
Similarly there is also an exact scaling relation for the third 
order cross correlation given by Yaglom\cite{yaglom}: 
$\la \De u_{r}(\De T_{r})^2 \ra =-4/3 \la N \ra r$. 
To demonstrate the quality of the inertial
range scaling of our DNS data, in Fig. 1 we present
 $-\la (\De u_{r})^3 \ra/r \la\ep\ra$ and
$-\la \De u_{r}(\De T_{r})^2 \ra /r \la N \ra$ as functions 
of $r$. It is seen that both quantities show a narrow 
inertial scaling range (less than one decade and they are also
displaced with respect to each other). On the other hand,
although DNS was carried out at moderate Reynolds numbers, 
the values of the dimensionless constants
agree well with theoretical predictions (4/5 and 4/3 for velocity and 
passive scalar, respectively). It was reported in
reference \cite{zhu} that these dimensionless constants were smaller
in the experimental measurements (for a circular jet
with ${\cal R}_\lambda = 220$ and the atmospheric surface layer with
${\cal R}_\lambda = 7200$) than the theoretical values.
In the inset, we show the passive scalar spectrum, $E_p(k)$, as a 
function of the wave number, $k$. The inertial range with 
a slope slightly less than $-5/3$\cite{obu} could be identified.
This smaller than $-5/3$ exponent was previously observed 
experimentally\cite{ze1}. A bump between the inertial and 
dissipation ranges perhaps is associated with 
the ``bottleneck effect'' \cite{lose}. 

In Fig. 2, we plot structure functions, $\la |\De T_r|^p \ra$, as
functions of $r$ for $p = 2,4,6$ and $8$. The power law inertial range 
can be identified for $0.2 \le r \le 1$. From data analysis
we notice that in fact the passive scalar increment displays a better 
scaling relation than the velocity increment\cite{cao}, and therefore
it is not necessary to invoke the Extended-Self-Similarity (ESS) 
technique\cite{benzi}. As a matter of fact we have 
tried ESS by plotting $\la |\De T_r|^p \ra$ as a function of
 $\la \De u_r(\De T_r)^2 \ra$. The error bar for the extracted scaling
exponents in the inertial range seems bigger than the result using 
the original structure functions. In inset (a), we 
show the local slope, $z_p(r) = d \log S_p(r)/d \log r$, as a function
of $r$ for the same $p$ in the structure function.
The local slope was calculated using a least square fit to a power law
for every 3 neighboring points. A flat region for each $p$ can be seen for 
$z_p(r)$, supporting the existence of inertial
scaling range. In inset (b), we present the 
two point correlation function of the passive scalar 
dissipation function, $\la N(x) N(x+r) \ra$, as a function of $r$.
Again a scaling relation for the dissipation correlation is found: 
$\la N(x)N(x+r) \ra \sim r^{-\mu}$ with $\mu \approx 0.25$. The intermittency
parameter $\mu$ agrees well with previous experimental results\cite{antonia}.

\bigskip
\psfig{file=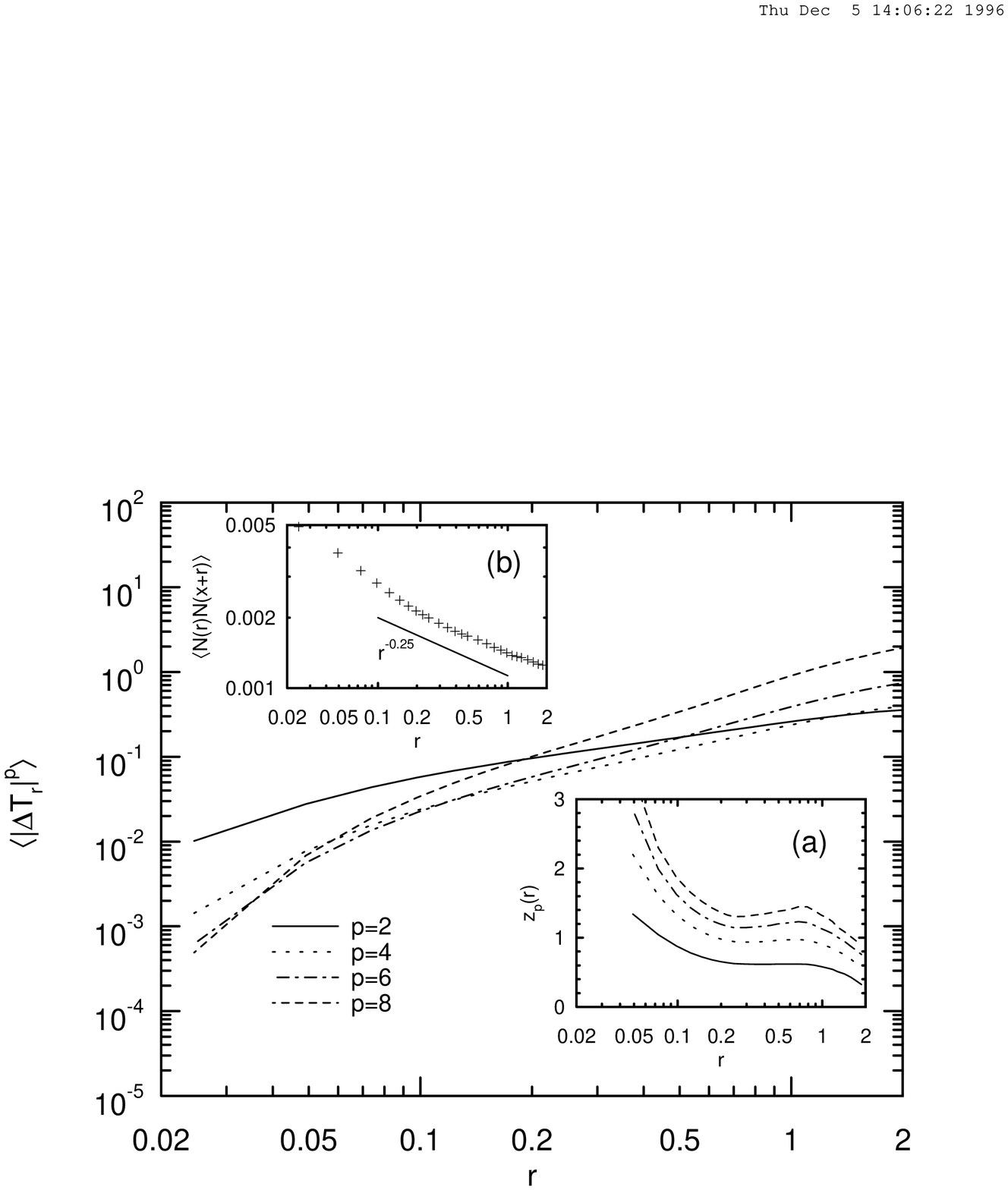,width=220pt}
\noindent
{\small FIG.~2.
The structure function $S_p(r)$ as a function of $r$.
In the inset (a) shows the local slope $z_p(r)$ as a function of $r$ and
in the inset (b), we present the
two-point correlation function of the passive scalar dissipation.}
\bigskip

In Fig. 3, we show the normalized probability density function (PDF),
$\la D \ra P(D)$, as a function of the normalized fluctuation dissipation:
$D/\la D \ra$, where $D$ is the dissipation function.
It is clear that both the velocity dissipation and the 
scalar dissipation fields are strongly intermittent\cite{sreeni-an}.
We have also compared the PDFs with the log-normal distribution by looking 
at the statistics of
 $w_D=(ln D-\la ln D \ra)/\la(ln D-\la ln D\ra)^2\ra^{1/2}$.
It is found that both fields are quite close to the log-normal values for low
order moments.
From the plot, we also note that the 
PDF of the scalar dissipation has a wider tail than that of the 
velocity dissipation, indicating that spatially
there are more large amplitude events in the passive scalar dissipation 
than the velocity dissipation. In the inset we show the flatness of 
the passive scalar increment and the longitudinal velocity increment 
as functions of the separation. It is seen that the flatness of the 
scalar increment is larger than that of the longitudinal 
velocity increment, except when $r$ is in the large scale region 
where both fields are essentially Gaussian and their flatnesses 
approach the Gaussian value of $3$. In particular, 
when $r$ is in the dissipation range, the 
flatness of the passive scalar increment is significantly larger than 
that of the velocity increment, in agreement with the PDFs. 

\bigskip
\psfig{file=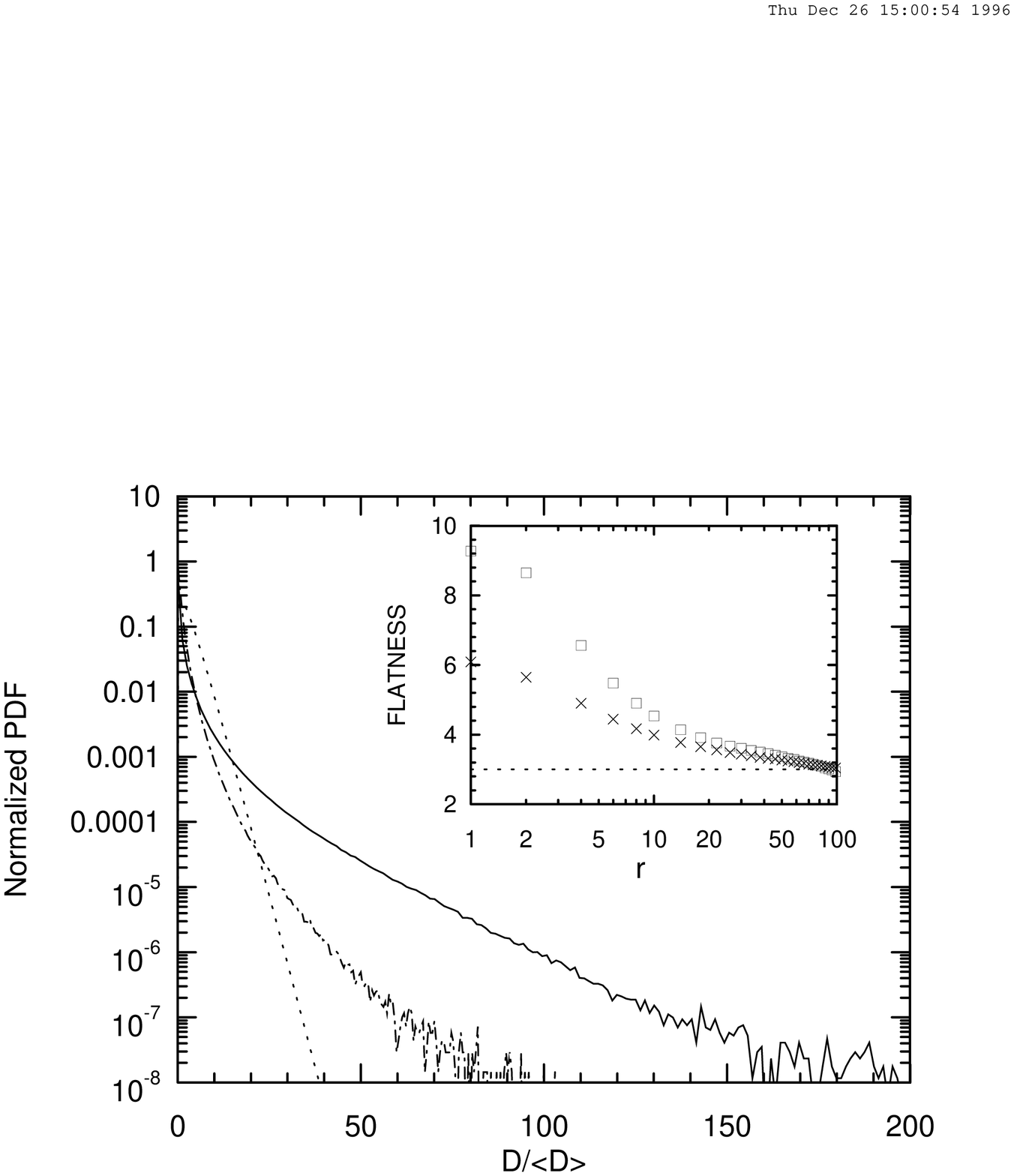,width=220pt}
\noindent
{\small FIG.~3. 
Normalized PDF, $\la D \ra P(D)$ as a function of 
the normalized dissipation. Here $D$ is the dissipation function (the solid
line is for the scalar dissipation and the dotted-dash line for the
velocity dissipation). The dotted line is for the scalar dissipation function 
with Gaussian statistics. In the inset we present the flatness of 
the scalar increment ($\Box$) and the velocity increment ($\times$)
as a function of $r$ (lattice unit).}
\bigskip

\psfig{file=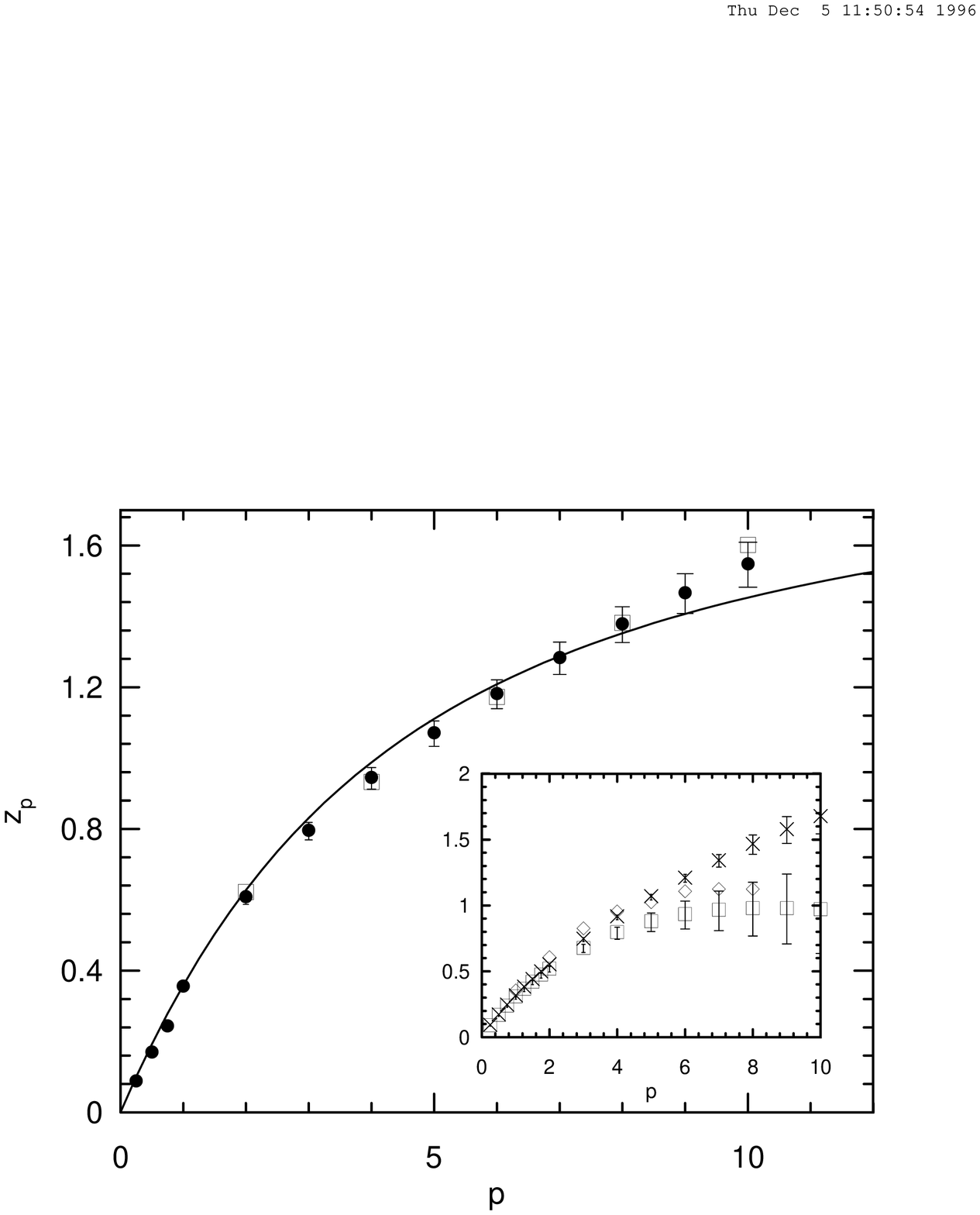,width=220pt}
\noindent
{\small FIG.~4.
The scaling exponent, $z_p$, as a function of $p$.
The $\bullet$ sign is for DNS and
the $\Box$ for the experiment data in \cite{antonia}. The solid line is for a
bi-variate log-Poisson model in \cite{cao-chen}. In the inset shows the
scaling exponents at two different time steps without using time averaging.
The data from \cite{meneveau} are represented by the $\diamond$ sign.}
\bigskip

The scaling exponent, $z_p$, as a function of the order 
index, $p$, is shown in Fig. 4. The $\bullet$ sign is for the result from 
DNS measurement and the $\Box$ sign for the result from the experiment by
Antonia {\em et al.}\cite{antonia}. The solid line is from a
bi-variate log-Poisson model in \cite{cao-chen} assuming that the
correlation coefficient between velocity field and the passive scalar field is 
zero. The DNS scaling exponents are obtained using the 
flat region in the the local slope plot as shown in the
inset (a) of Fig. 2. Some typical scaling exponents and 
the corresponding errors are listed here: $z_{0.5} = 0.165 \pm 0.007,
z_1 = 0.354 \pm 0.012, z_2 = 0.606 \pm 0.019, z_3 = 0.794 \pm 0.025,
z_4 = 0.943 \pm 0.030, z_6 = 1.180 \pm 0.041, z_8 = 1.376 \pm 0.050$ and
$z_{10} = 1.546 \pm 0.063$. It is noticed that the second order scaling
exponent deviates from Obukhov's $2/3$ law by 15\%.
This result agrees well with 
previous experimental results\cite{antonia,meneveau,ze1}. The fourth order 
scaling exponent in DNS also agrees quite well with the theoretical 
prediction: $z_4 = 2 z_2 - \mu$ in \cite{rhk1}, a result primarily
for the white and Gaussian velocity. In general, all scaling 
exponents from DNS coincide well with the phenomenological 
theory\cite{cao-chen} and the real-life experiment\cite{antonia}.

We point out that the DNS results shown in Fig.4 were obtained 
using a spatial averaging over the whole simulation domain ($512^3$)
and a long-time averaging ($\sim 10$ large-eddy turnover times). In fact, 
the scaling exponent for each single time frame displays intense fluctuation 
with time, much larger than the fluctuation of the exponents for the velocity 
increment\cite{cao} using a spatial averaging of the same size. To 
demonstrate this fluctuation, in the inset of Fig. 4, two typical scaling 
exponents without time averaging are shown. The time difference between 
two frames is about 0.2 large-eddy turnover time. Although the curves 
for $p\le4$ agree quite well, they are qualitatively different for larger 
$p$. A large variation of the corresponding PDFs with time has 
also been observed. One set of the 
scaling exponents ($\Box$) clearly establishes a saturation when $p \ge 7$, 
indicating a possible upper bound for the most intensive
events. The saturation of scaling exponents for the
passive scalar has been observed in \cite{meneveau} (shown also in the inset 
by the $\diamond$ sign) and \cite{borue}. 
The scaling exponents in another group ($\times$) are
larger than the time-averaged values in Fig. 4 for $p \geq 5$.
Such strong fluctuations could be a reason for the saturation displayed
in \cite{meneveau} where the results were obtained based on 1-D cuts 
through the field and may not capture the strongest events with a 
limited sample. On the other hand, the experimental scaling exponents were 
obtained through the RSH and it can not be ruled out that 
the saturation could be due to shortcomings of the RSH at high values of $p$.

To understand the physics behind the large time fluctuation of the scaling 
exponent, we have studied the intermittency structures of the 
passive scalar field, in particular the large amplitude events of 
the dissipation function and the scalar derivative. The latter is 
directly associated with the scalar increment: 
$\De T_r = \int_0^{r} \partial (T/\partial x) dx$. Using visualization
technique, we find that with increasing of amplitude, the iso-surface of the
scalar dissipation changes from fragment-like to sheet-like (shown in 
Fig. 5). The dynamical evolution of the sheet-like structures, such as 
formation and annihilation, clearly plays a key role in the time dependence 
of the scaling exponents, simply because the large amplitude
events are the predominate effects on the high order scalar increments and
their exponents. From the structure stability point of 
view, we suspect that the sheet-like structure are more unstable to 
perturbations, such as  forcing, than the filament-like 
structure\cite{she}, which is a characteristic vortex
structure for large amplitude events in three-D turbulence. 
This agrees with the observation that 
the velocity scaling exponent seems to be less time dependent. 
The instability of the sheet-like structures 
possibly causes the annihilation of the very high intermittency structures
at a specific time, leading to a saturation of the scaling exponents. 
In addition, we know from Fig. 3 that spatially 
there are more high amplitude events in the scalar field than in 
the velocity field, indicating that the generation and the annihilation of the 
sheet-like structure is more frequent and causes intense fluctuations 
in the statistics of the scalar field. 

\vspace{1.1in}
\psfig{file=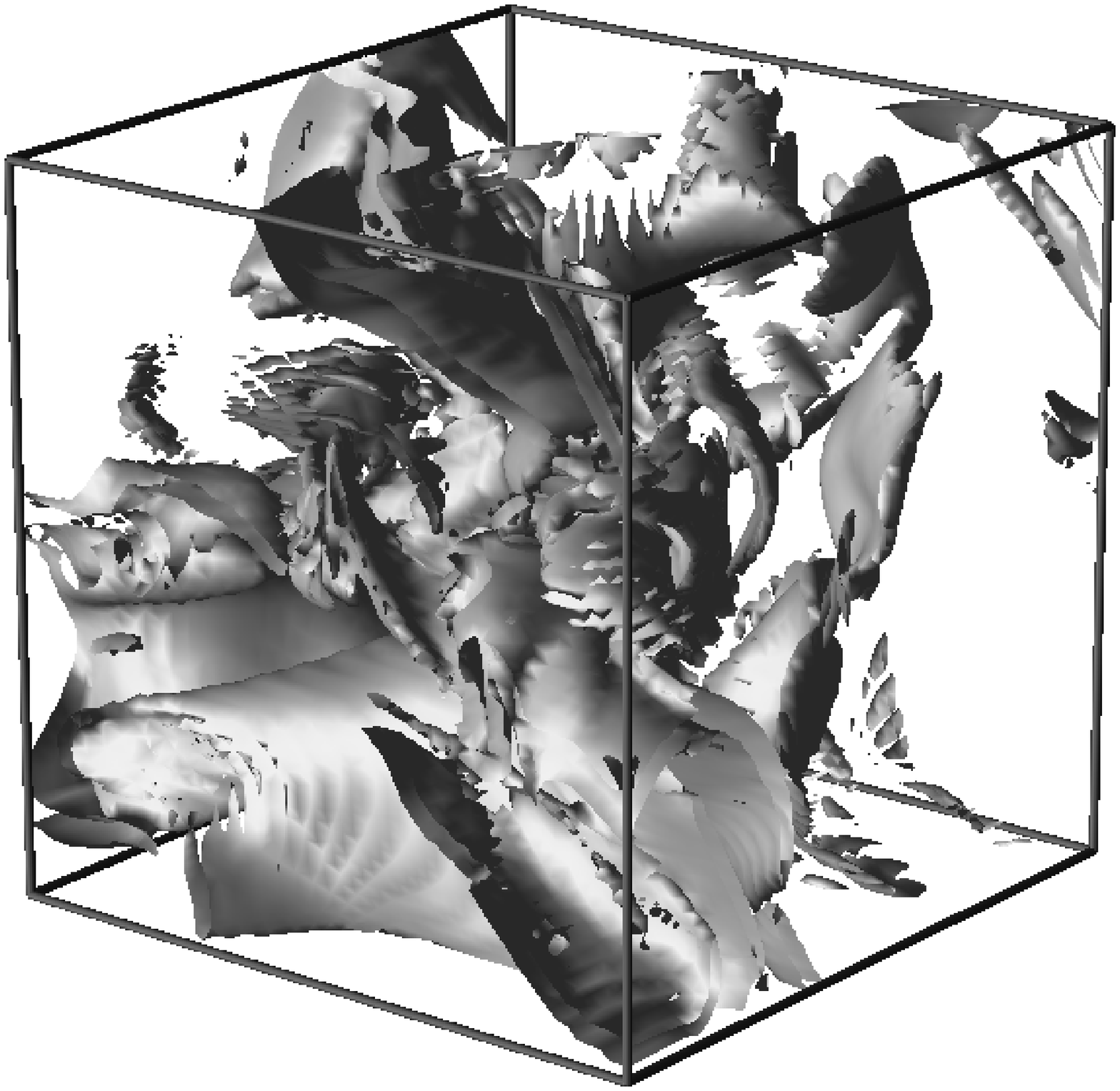,width=160pt}
\noindent
{\small FIG.~5.
Iso-surface of the scalar dissipation at the value
$N = 10 \la N \ra$. The resolution of 
the plot is $64^3$ and the data shown represent the
neighborhood of the overall maximum of scalar dissipation.}
\bigskip

The current finding of the time dependence of the scaling exponents might 
connect with the idea of the non-universality of scaling exponents for 
the passive scalar system\cite{siggia} in the sense that the detailed 
intermittency structure strongly affects the exponents.  
On the other hand, so far little is known about how the 
velocity stretching affects the dynamical evolution of the 
sheet-like structure in the scalar field. We feel that the fundamental 
physics of the scaling dynamics for the passive scalar turbulence is still 
missing and a more detailed and careful study is much needed.

We thank R. Kraichnan, C. Meneveau, M. Nelkin, Z.-S. She and
K. R. Sreenivasan for useful discussions. Numerical simulation was carried 
out at the Advanced Computing Laboratory at Los Alamos National Laboratory.

\end{multicols}
\end{document}